\documentclass[prd,aps,twocolumn,showpacs,preprintnumbers,amsmath,amssymb,nofootinbib]{revtex4}
\usepackage[dvips]{graphicx}
\usepackage{subfig}
\usepackage{dcolumn}
\usepackage{ulem}

\begin{document}

\title{The signature of the scattering between dark sectors in large scale cosmic microwave background anisotropies }

\author{Xiao-Dong Xu, Bin Wang}   \affiliation{INPAC and Department of Physics, Shanghai Jiao Tong University, 200240 Shanghai, China}
\author{Elcio Abdalla}  \affiliation{Instituto de Fisica, Universidade de Sao Paulo, CP 66318, 05315-970, Sao Paulo, Brazil}

\begin{abstract}
We  study the interaction between dark sectors by
considering the momentum transfer caused by the
dark matter scattering elastically within the
dark energy fluid. Describing the dark scattering
analogy to the Thomson scattering which couples
baryons and photons, we examine the impact of the
dark scattering in CMB observations. Performing
global fitting with the latest observational
data, we find that for a dark energy equation of
state $w<-1$, the CMB gives tight constraints on
dark matter-dark energy elastic scattering.
Assuming a dark matter particle of proton mass,
we derive an elastic scattering cross section of
$\sigma_D < 3.295 \times 10^{-10} \sigma_T$ where
$\sigma_T$ is the cross section of Thomson
scattering. For $w>-1$, however, the constraints
are poor. For $w=-1$, $\sigma_D$ can formally
take any value.

\end{abstract}

 \maketitle

\section{Introduction}

There has been convincing evidence indicating that our universe is expanding at an increasing rate and it is mainly composed of
dark matter (DM) and  dark energy (DE) at the present moment. The simplest explanation of DE is the cosmological constant, that fits observations
very well. However, the observed value of the cosmological constant  falls far below the value predicted by any sensible quantum field theory.
This is the so called cosmological constant problem. Furthermore using the cosmological constant to explain the DE
unavoidably leads to the coincidence problem, namely, why are the vacuum and matter energy densities of precisely the same order today\cite{Weinberg00}\cite{Carroll01}?

Considering that DE and DM contribute
significant fractions of the contents of the universe, it is
natural, in the framework of field theory, to consider the
interaction between them. The covariant description of the interaction between DE and DM is given by \cite{Kodama}
$\nabla_{\mu}T^{\mu\nu}_{(\lambda)} = Q^{\nu}_{(\lambda)}$, where $Q^{\nu}_{(\lambda)}$ is a four vector governing the energy momentum
transfer between DE and DM.  The subindex $\lambda$ refers to DM or DE component respectively.
For the whole system, the energy and momentum are conserved, and the transfer vector satisfies $\sum_{\lambda}Q^{\nu}_{(\lambda)} = 0$.
The four vector $Q^{\nu}_{(\lambda)}$ describing the interaction between DE and DM can be phenomenologically decomposed
into two parts with respect to a given observer $\lambda'$  with four velocity $U^{\mu}_{(\lambda')}$ \cite{He.prd83}, namely,
$Q^{\mu}_{(\lambda)} = Q_{(\lambda)}|_{\lambda'}U^{\mu}_{(\lambda')} + F^{\mu}_{(\lambda)}|_{\lambda'}$, where
$Q_{(\lambda)}|_{\lambda'} = -U_{(\lambda')\nu}Q^{\nu}_{(\lambda)}$ is the energy transfer rate of the $\lambda$ component
observed by  the observer situated on the $\lambda'$ component. $F^{\mu}_{(\lambda)}|_{\lambda'} = h^{\mu}_{(\lambda')\nu}Q^{\nu}_{(\lambda)}$ is the
corresponding momentum transfer between the two components observed by the observer in the $\lambda'$ frame.

The possibility that DE and DM interact with energy exchange between them has been studied extensively \cite{He.prd83}-\cite{Baldi.MNRAS11}.
It has been shown that the energy transfer between DE and DM can provide a mechanism to alleviate the coincidence problem \cite{He.prd83}-\cite{He.prd80}. Complementary observational signatures of the energy exchange between DE and DM have been obtained from the cosmic expansion history by using the WMAP, SNIa, BAO and SDSS data etc\cite{He.prd83}\cite{He.prd80}-\cite{Xu:PLB11} and the growth of cosmic structure \cite{He.jcap07}-\cite{Baldi.MNRAS11}.

Besides the energy transfer between DE and DM, their interaction may also impart a transfer of momentum.
Considering the extremely low DE density, and the nonrelativistic velocities of DM motions, elastic scattering appears naturally
between the DM and DE fluid.  We do not need to restrict to a particular DE model since the macroscopic physics is independent of the microphysics in the
scattering process. The DM scattering elastically within the DE fluid results in the momentum transfer between them.
The implications of the elastic scattering between DE and DM was explored in \cite{Simpson}. It was found that the growth of structure
was suppressed by a drag term arising from elastic scattering between dark sectors. It is of great interest to extend the
study in \cite{Simpson} to investigate the cosmological signal of the elastic scattering between the DM and DE fluid. In this work we will explore the
implications of the momentum transfer between dark sectors in the cosmic background radiation and  constrain the scattering cross section
between DE and DM by using the WMAP observation and the joint analysis of CMB, SNIa and BAO data.

\section{Perturbation formalism with Dark Scattering}

In this section, we construct the first order perturbation theory when there is elastic scattering between the DM and the DE fluid.
The perturbed space-time at first order reads
\begin{equation}
\begin{split}
ds^2 =& a^2 [ -(1+2\psi)d\tau^2 + 2\partial_iBd\tau dx^i \\
&+ (1+2\phi)\delta_{ij}dx^idx^j + D_{ij}Edx^idx^j ]
\end{split}
\end{equation}
where $a$ is the cosmic scale factor, $\psi, \phi, B, E$ are the scalar metric perturbations and
$D_{ij} \equiv (\partial_i\partial_j - \frac{1}{3}\delta_{ij}\nabla^2)$. Choosing the Newtonian gauge, we take $B=E=0$. In the background
the energy conservations of DM and DE are described by
\begin{align}
& \rho_c' + 3\mathcal{H}\rho_c = aQ_c, \\
& \rho_d' + 3\mathcal{H}(1+w)\rho_d = aQ_d,
\end{align}
where the subscripts `$c$' and `$d$' refer to DM
and DE respectively. A prime denotes the
derivative with respect to the conformal time
$\tau$.  $w$ represents the dark energy equation
of state and $\mathcal{H} \equiv \frac{a'}{a}$ is
the Hubble parameter. Here $Q_c$ and $Q_d$ are
the energy exchange between DE and DM observed in
background. If there is no energy transfer
between the dark sectors, $Q_c=Q_d=0$, the energy
densities of DM and DE are separately conserved.

In the Fourier space, the covariant form of the perturbed energy-momentum transfer between DE and DM can be expressed as \cite{He.plb671}\cite{He.jcap07}
\begin{align}
\begin{split}
&\delta'_{\lambda} + 3\mathcal{H}(\frac{\delta p_{\lambda}}{\delta\rho_{\lambda}} - w_{\lambda})\delta_{\lambda} + (1+w_{\lambda})kv_{\lambda} \\
&= -3(1+w_{\lambda})\phi' + (2\psi - \delta_{\lambda})\frac{a^2Q^0_{\lambda}}{\rho_{\lambda}} + \frac{a^2\delta Q^0_{\lambda}}{\rho_{\lambda}},
\label{delta}\end{split} \\
\begin{split}
&v_{\lambda}' + \mathcal{H}(1-3w_{\lambda})v_{\lambda} - \frac{k}{1+w_{\lambda}} \frac{\delta p_{\lambda}}{\delta \rho_{\lambda}}\delta_{\lambda} \\
&= -\frac{w_{\lambda}'}{1+w_{\lambda}}v_{\lambda}
+ k\psi - \frac{a^2
Q^0_{\lambda}}{\rho_{\lambda}}v_{\lambda} +
\frac{a^2\delta
Q_{p\lambda}}{(1+w_{\lambda})\rho_{\lambda}},
\label{v}\end{split}
\end{align}
where $\delta_{\lambda} \equiv
\delta\rho_{\lambda}/\rho_{\lambda}$ and
$v_{\lambda}$ is the peculiar velocity. $\delta
Q_{p\lambda}$ is the potential of the spatial
part of the perturbed coupling vector, $\delta
Q^i_{\lambda}$. As discussed above, $\delta
Q_{p\lambda}$ can be decomposed into two parts,
$\delta Q_{p\lambda} =
Q_{\lambda}|_{\lambda'}v_{\lambda'} +
F_{\lambda}|_{\lambda'}$. $v_{\lambda'}$ is the
velocity  of the observer situated in the
$\lambda'$ component and
$F_{\lambda}|_{\lambda'}$ is the external
non-gravitational force density on $\lambda$
component observed by the observer sitting in the
$\lambda'$ component. In the following
discussion, we will assume that the energy
transfer vanishes in the background and only
concentrate on the momentum transfer between dark
sectors. Therefore, in the background frame,
$Q_{\lambda}|_{(back)} = 0$, and only
$F_{\lambda}|_{(back)}$ contributes to the energy
momentum transfer.  $F_{\lambda}|_{(back)}$ is
simplified as $F_{\lambda}$ below.

In the absence of a fundamental theory, we can only conceive a phenomenological description of DM scattering elastically within the DE fluid.
Before we do so, we recall the description of the coupling between baryon and photon fluid. The perturbation equations of baryon read
\begin{align}
&\delta_b' + 3C_s^2\mathcal{H}\delta_b + 3\phi' = -kv_b, \label{deltab}\\
&v_b' = -\mathcal{H}v_b + kC_s^2\delta_b + k\psi + an_e\sigma_T \frac{4\rho_{\gamma}}{3\rho_b}(v_{\gamma} - v_b), \label{vb}
\end{align}
where the subscript `$b$' represents baryon and
`$\gamma$' stands for photon.  $C_s$ is the sound
speed. The last term in (\ref{vb}) arises from
Thomson scattering. Comparing (\ref{deltab}) and
(\ref{vb}) with (\ref{delta}) and (\ref{v}), we
found that $Q^0_b = 0$, $\delta Q^0_b = 0$ and
\begin{equation}
F_b = n_e\sigma_T(1+w_{\gamma})\rho_{\gamma}(v_{\gamma} - v_b)/a.
\end{equation}
Here $\sigma_T$ is the cross section between baryon and photon. Using the analogy, we can choose the interaction term of DM due to the elastic
scattering in the DE fluid
\begin{equation}
F_c = \frac{1}{a}(1+w)\rho_dn_c\sigma_D(v_d-v_c),
\label{Fc}\end{equation} where $\sigma_D$
represents the unknown cross section of the
elastic scattering between DM and DE, $n_c =
\rho_c/m_c$ is the number density of DM
particles. Conservation of momentum leads to a
similar term arising for DE.  It is interesting
that $(1+w)$ appears in the interaction term. We
can see that this is a general requirement for
momentum transfer. Given $\lambda$ refers to dark
energy in (\ref{v}), $F_d = -F_c$ is divided by
$(1+w)$, which must vanishes when $w$ approaches
$-1$ in order to avoid singularities. However,
this does not imply that momentum transfer has no
impact on the dark sector perturbations for $w
\sim -1$. Since we do not have a good estimation
of the mass of DM particles, it is convenient to
define $\Sigma \equiv \sigma_D/m_c$ and express
the momentum transfer as
\begin{equation}
F_c = \frac{1}{a}(1+w)\rho_d\rho_c\Sigma(v_d-v_c).
\label{Fd}\end{equation}
This is the ansatz of the dark scattering adopted in \cite{Simpson}. In the study of the scattering between baryon and dark fluid,
the same interaction term for the dark fluid was adopted in \cite{Aviles}. It would be fair to say that we do not
know the microphysics in quantizing the DE so that we do not have the exact definition of the cross section of the elastic scattering between DM and DE.
In \cite{Simpson} it was argued that the analysis of the macroscopic behavior are largely independent of the microphysics involved and the
bound on the DM and DE cross section was first derived from the impact incurred on the growth of large scale structure. In this work we are going to
investigate the signature of this phenomenological term on the DM and DE elastic scattering in the CMB observations. For the sake of simplicity,
we will set $\Sigma$ to be constant in the following.

Inserting (\ref{Fc}) into the general equations (\ref{delta}) and (\ref{v}), we get, in the Newtonian gauge,
the evolution equations of the perturbations to DM and DE
\begin{align}
\delta_c' =& -kv_c - 3\phi', \\
v_c' =& -\mathcal{H}v_c + k\psi + a(1+w)\rho_d\Sigma(v_d-v_c), \\
\delta_d' =& -3\mathcal{H}(\frac{\delta p_d}{\delta\rho_d} - w)\delta_d - (1+w)kv_d \nonumber\\
&- 3(1+w)\phi', \\
v_d' =& \frac{k}{1+w}\frac{\delta p_d}{\delta\rho_d}\delta_d - \mathcal{H}(1-3w)v_d - \frac{w'}{1+w}v_d + k\psi \nonumber\\
&+ a\rho_c\Sigma(v_c-v_d).
\end{align}
We choose the Newtonian gauge where the Thomson
scattering is well established. Now we rewrite
the equations in the gauge invariant quantities
so that our calculations below do not depend on
the specific gauge choice. Constructing the gauge
invariant quantities \cite{He.prd80}
\begin{align*}
& \Psi = \psi - \frac{\mathcal{H}}{k}(B+\frac{E'}{2k}) - \frac{1}{k}(B' + \frac{E''}{2k})\quad, \\
& \Phi = \phi + \frac{E}{6} - \frac{\mathcal{H}}{k}(B+\frac{E'}{2k})\quad , \\
& D_{g\lambda} = \delta_{\lambda} - \frac{\rho_{\lambda}'}{\rho_{\lambda}\mathcal{H}}(\phi+\frac{E}{6})\quad , \quad V_{\lambda} = v_{\lambda} - \frac{E'}{2k}\quad , \\
& \delta Q^{0I}_{\lambda} = \delta Q^0_{\lambda} - \frac{{Q_{\lambda}^0}'}{\mathcal{H}}(\phi+\frac{E}{6}) + Q^0_{\lambda}\bigg[\frac{1}{\mathcal{H}}(\phi+\frac{E}{6})\bigg]', \\
& \delta Q^I_{p\lambda} = \delta Q_{p\lambda} - Q^0_{\lambda}\frac{E'}{2k}\quad ,
\end{align*}
we obtain the gauge-invariant perturbation equations
\begin{align}
D_c' =& -kV_c \label{Dc}\quad ,\\
V_c' =& -\mathcal{H}V_c + k\Psi + a(1+w)\rho_d\Sigma(V_d-V_c) \label{Vc}\quad ,\\
D_d' =& -k(1+w)V_d - 3\mathcal{H}(C_e^2-w)D_d \nonumber\\
     &- 9\mathcal{H}^2(C_e^2-C_a^2)(1+w)\frac{V_d}{k} \nonumber\\
     &+ \{3w' + 9\mathcal{H}(C_e^2-w)(1+w)\} \Phi \label{Dd}\quad ,\\
V_d' =& -\frac{w'}{1+w}V_d - \mathcal{H}(1-3w)V_d + kC_e^2\frac{D_d}{1+w} \nonumber\\
     &+ 3\mathcal{H}(C_e^2-C_a^2)V_d - 3kC_e^2\Phi + k\Psi \nonumber\\
     &+ a\rho_c\Sigma(V_c-V_d)\quad . \label{Vd}
\end{align}
We have employed
\begin{equation}
\frac{\delta p_d}{\rho_d} = C_e^2\delta_d - (C_e^2-C_a^2)\frac{\rho_d'}{\rho_d}
\frac{V_d+B}{k}\quad ,
\end{equation}
where $C_e^2$ is the effective sound speed of  DE
and $C_a^2 = \frac{p_d'}{\rho_d'}$ is the
adiabatic sound speed of DE. We assume $C_e^2 =
1$ in our calculation below.  The elastic
scattering between DM and DE does not enter the
equations governing the perturbations of energy
densities, but affects the perturbations in the
peculiar velocities.

\subsection{Stability of the perturbation}

It is known that DE perturbation is gravitationally unstable when its equation of state crosses $-1$ \cite{Vikman}-\cite{Doran.prd72}.
Such an instability happens regardless of whether there is interaction between DE and DM or not. To avoid this instability we restrict $1+w$ to be either
positive or negative in the following discussion.

When there is energy transfer between DE and DM, it was observed that the stability of the perturbation depends on the form of the interaction
and the equation of state of DE \cite{He.prd83} \cite{Xu:PLB11}-\cite{Maartens.jcap07}.

\begin{figure}[hptb]
\subfloat[]{
    \includegraphics[width=0.5\textwidth]{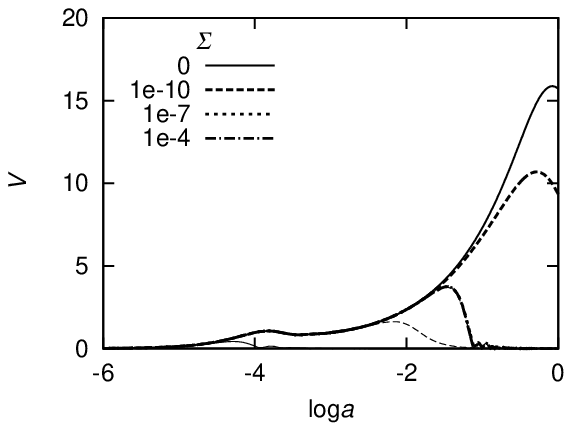}
    \label{v-cq}
}\\
\subfloat[]{
    \includegraphics[width=0.5\textwidth]{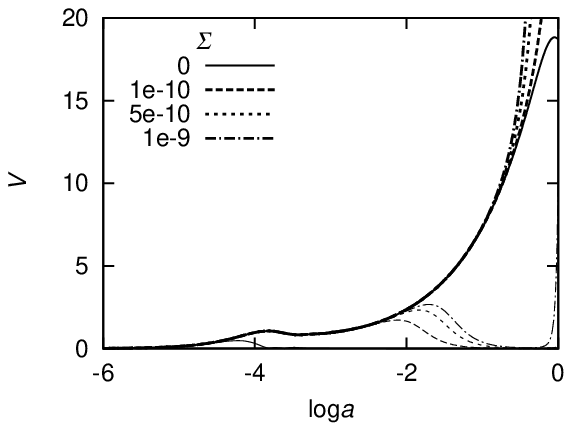}
    \label{v-cp}
}\\
\subfloat[]{
    \includegraphics[width=0.5\textwidth]{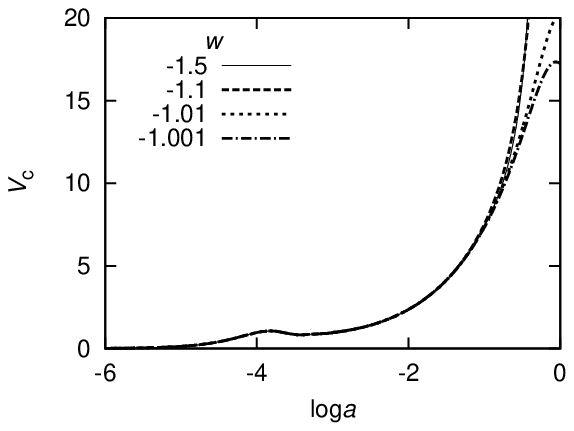}
    \label{v-w}
} \caption{The evolutions of peculiar velocities
for $k = 0.1 \textrm{Mpc}^{-1}$.  $\Sigma$ is in
the unit of $\sigma_T / m_p$ where $\sigma_T$ is
the cross section of Thomson scattering and $m_p$
is proton mass. In (a) and (b), thick lines stand
for $V_c$ and thin lines represent $V_d$. In (a)
we choose $w = -0.8$ and in (b) $w = -1.5$. In
(c) we show the behavior of peculiar velocity
evolution of $V_c$ when $w$ approaches $-1$ from
below when we fix $\Sigma=1 \times
10^{-9}\sigma_T/m_p$. \label{V-c}}
\end{figure}

It is of interest to examine the stability of the perturbation when there is momentum transfer between DM and DE.
We find that for the DE equation of state satisfying $1+w>0$, there is no instability in the perturbation caused by the scattering between dark sectors.
The dark scattering acts as a new drag term in the perturbation equations.
In the early period, when $a\rho_c$ dominated and $a\rho_d$ was small, we observe that $V_c$ evolved freely while $V_d$ was forced to follow
along with $V_c$. With the decrease of $a\rho_c$, $V_d$ gradually departed from $V_c$ and decayed. In the late time, when $a(1+w)\rho_d$
became important, the dark scattering will drag $V_c$ to $V_d$. The effect of the scattering between DE and DM on the behavior of the evolution of
$V_c, V_d$ can be clearly seen in Fig.\ref{v-cq}.

In the case when $(1+w)<0$, the sign of the last term in (\ref{Vc}) is flipped and the drag force turns into propulsion. This will cause the blow up
of the peculiar velocities of DM and DE, which are shown in Fig.\ref{v-cp}. To understand the blow up of $V_c$, we can simply look at
$V_c' \sim a(1+w)\rho_d\Sigma(V_d-V_c)$. Considering $|V_d|<|V_c|$ in the late time, $w<-1$ will cause the exponential growth of $V_c$.
In turn the exponential growth of $V_c$ will also lead the blow up of $V_d$ due to the last term in (\ref{Vd}). The instability due to dark scattering is driven by the term $a(1+w)\rho_d\Sigma(V_d-V_c)$, which is suppressed when $w$ is close to $-1$. This can be seen clearly in Fig.\ref{v-w}.

\subsection{Tight coupling approximation}

Since we do not consider the energy transfer between DE and DM, the DM energy density $\rho_c$ scales as $a^{-3}$
and the combined term $a\rho_c\Sigma$ scales as $a^{-2}$. The overwhelmingly large value of $a\rho_c\Sigma$ in the early time becomes the bottleneck in the computation which severely limits the step size and the speed in the numerical calculation.
It precludes a straightforward numerical integration at early times:
tiny errors in the propagation of $V_c$ and $V_d$ will lead to serious errors to the force.
This problem is similar to that encountered when the baryon and photon fluids were strongly coupled via Thompson scattering at early times
\cite{Doran.jcap05}.  One resorted to the so called tight coupling approximation to improve the numerical computation in CMB. Here we will borrow
this method and eliminate the term of order  $a\rho_c$ and $a\rho_d$ from the evolution equation.

Subtracting (\ref{Vd}) from (\ref{Vc}) and doing the rearrangement, we have
\begin{equation}
\begin{split}
& a[(1+w)\rho_d+\rho_c]\Sigma S = -S' + \mathcal{H}V_c \\
&+ \bigg(-\frac{w'}{1+w} + \mathcal{H}T\bigg)V_d + kC_e^2\frac{D_d}{1+w} - 3kC_e^2\Phi,
\end{split}\label{slip}
\end{equation}
where $S \equiv V_d-V_c$ and $T \equiv -(1-3w) + 3(C_e^2-C_a^2)$. Taking derivative with respect to the conformal time, we have
\begin{equation}
\begin{split}
& a'[(1+w)\rho_d+\rho_c]\Sigma S + a[\{(1+w)\rho_d\}' + \rho_c']\Sigma S \\
&+ a[(1+w)\rho_d+\rho_c]\Sigma S' \\
&= -S'' + \mathcal{H}'V_c + \mathcal{H}V_c' + \bigg(-\frac{w''}{1+w} + \frac{w'^2}{(1+w)^2} \\
&+ \mathcal{H}'T + \mathcal{H}T'\bigg)V_d + \bigg(-\frac{w'}{1+w} + \mathcal{H}T\bigg)V_d' \\
&- kC_e^2\frac{w'}{(1+w)^2}D_d + kC_e^2\frac{D_d'}{1+w} - 3kC_e^2\Phi'.
\end{split}\label{dslip}
\end{equation}
The derivatives appearing above can be written as
\begin{align*}
& \rho_c' = -3\mathcal{H}\rho_c\; , \quad \rho_d' = -3\mathcal{H}(1+w)\rho_d\; , \quad a' = a\mathcal{H}\quad .
\end{align*}
Inserting these expressions together with (\ref{Vc}) and ({\ref{Vd}}) in  (\ref{dslip}), it becomes
\begin{equation}
\begin{split}
& a[(1+w)\rho_d+\rho_c]\Sigma S' \\
=& -S'' - [(1+w)\rho_d - \rho_c]\frac{w'}{1+w}a\Sigma S \\
+& \{3(1+w)^2\rho_d - (T-2)\rho_c\}a\mathcal{H}\Sigma S + \mathcal{H}'V_c \\
-& \mathcal{H}^2V_c + \bigg(-\frac{w'}{1+w} + \mathcal{H}T\bigg)^2V_d \\
+& \bigg(-\frac{w''}{1+w} + \frac{w'^2}{(1+w)^2} + \mathcal{H}'T + \mathcal{H}T'\bigg)V_d \\
+& kC_e^2\frac{D_d'}{1+w} + \bigg(-2\frac{w'}{1+w} + \mathcal{H}T\bigg)kC_e^2\frac{D_d}{1+w} \\
+& 3kC_e^2\frac{w'}{1+w}\Phi - 3kC_e^2\mathcal{H}T\Phi - 3kC_e^2\Phi' \\
+& \bigg\{-\frac{w'}{1+w} + \mathcal{H}(1+T)\bigg\}k\Psi.
\end{split}\label{dslip2}
\end{equation}
Inserting (\ref{dslip2}) into (\ref{slip}) and in turn using in (\ref{Vc}) and (\ref{Vd}), we obtain the evolution equations for $V_c$ and $V_d$,
\begin{align}
\begin{split}
V_c' &= -\mathcal{H}V_c + k\Psi + \frac{1+w}{1+w+r}\bigg\{-S' - 3kC_e^2\Phi \\
&+ \mathcal{H}V_c + \bigg(-\frac{w'}{1+w} + \mathcal{H}T\bigg)V_d + kC_e^2\frac{D_d}{1+w} \bigg\},
\label{tightVc}\end{split} \\
\begin{split}
V_d' &= \bigg(-\frac{w'}{1+w} + \mathcal{H}T\bigg)V_d + kC_e^2\frac{D_d}{1+w} - 3kC_e^2\Phi \\
&+ k\Psi - \frac{r}{1+w+r}\bigg\{-S' - 3kC_e^2\Phi + \mathcal{H}V_c \\
&+ \bigg(-\frac{w'}{1+w} + \mathcal{H}T\bigg)V_d + kC_e^2\frac{D_d}{1+w} \bigg\},
\label{tightVd}\end{split}
\end{align}
where $r \equiv \rho_c/\rho_d$. There is no
longer any term proportional to $a\rho_c$ or
$a\rho_d$ in the above formulae. Up to now, we
have not yet made any approximation. To finish
the tight coupling approximation, we drop the
term $S''$ in (\ref{dslip2}) as has been done in
\cite{Doran.jcap05}. The second derivative of $S$
is the only unknown term during the numerical
integration of perturbations and  can be
neglected in the expression of $S'$ when DE and
DM are strongly coupled.  In the
tight coupling approximation we adopt here,
considering the DE density to be sufficiently low at
early times, $V_c$ can evolve along (\ref{Vc})
while for $V_d$ we need to employ the
approximated equation (\ref{tightVd}), in which
$S'$ is evaluated using (\ref{dslip2}) and $S''$
therein is neglected. This strategy can lead to
the efficiency in computation.

\section{Impact of the Dark Scattering on the CMB Power Spectrum}

With the perturbation equations at hand, we can study the influence of dark scattering on the CMB radiation.
We have modified the public available code CMBEASY \cite{easy} to account for the elastic scattering between DE and DM.
In our theoretical computation, we allow the equation of state of DE to be constant and time-dependent in the range  $w>-1$ and $w<-1$ respectively.
To see the effect of the scattering in CMB, a wide range of the strength of dark scattering is explored for the selected DE equation of state.

\begin{figure}[hptb]
\subfloat[]{
    \includegraphics[width=0.5\textwidth]{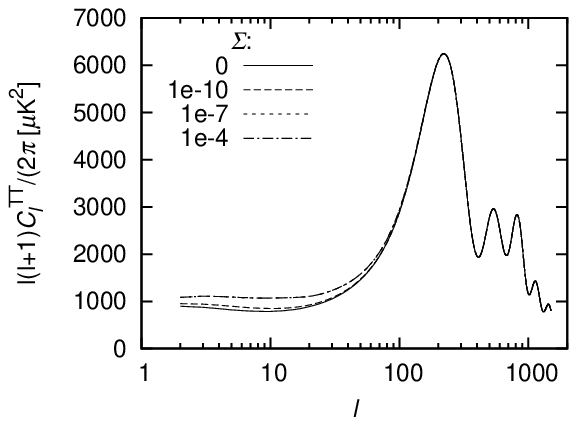}
}\\
\subfloat[]{
    \includegraphics[width=0.5\textwidth]{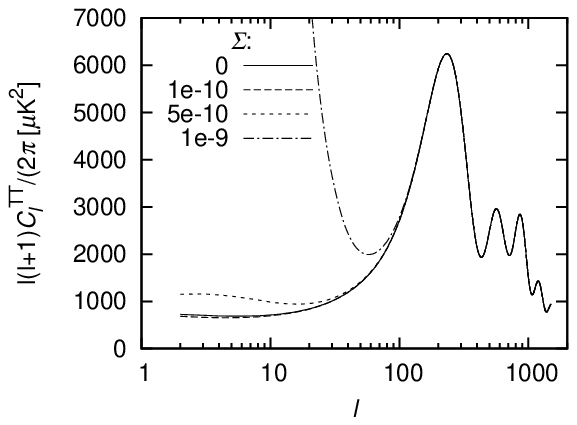}
    \label{cl-cp}
}
\caption{CMB angular power spectrum when there is dark scattering between DE and DM. $\Sigma$ is in the unit of $\sigma_T / m_p$. In (a) we choose constant DE equation of state $w = -0.8$.
In (b) the constant DE equation of state is fixed as $w = -1.5$.}
\label{cl-c}
\end{figure}

\begin{figure}[hptb]
\subfloat[]{
    \includegraphics[width=0.5\textwidth]{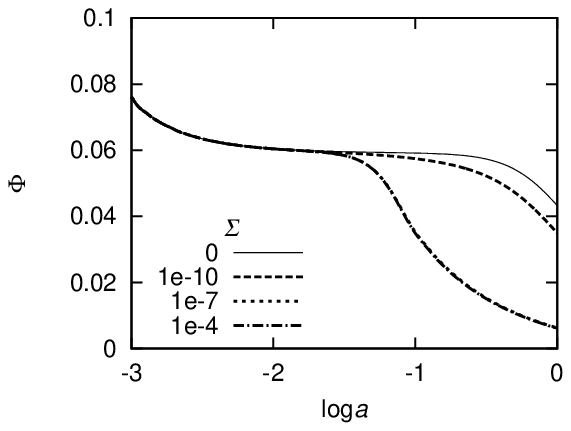}
}\\
\subfloat[]{
    \includegraphics[width=0.5\textwidth]{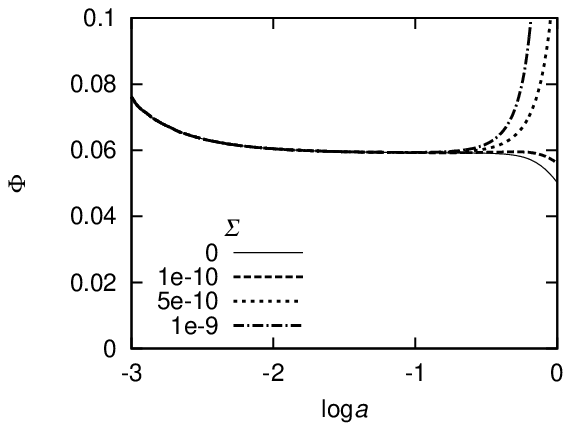}
    \label{phi-cp}
}
\caption{The evolution of gravitational potentials when $k = 0.1 \textrm{Mpc}^{-1}$. In (a) we choose $w = -0.8$ and in (b) we select $w = -1.5$.}
\label{phi-c}
\end{figure}

When $w$ is very close to $-1$, the momentum transfer due to the elastic scattering between DM and DE does not lead to a clearly discernible change on the CMB power spectrum. The dark scattering effect can obviously show up once the equation of state of DE deviates more from $-1$.
For a constant equation of state of DE, the dark scattering impact on CMB can be seen in Fig.\ref{cl-c}.  When the constant equation of state of DE $w>-1$,
we find that with the increase of the strength of the scattering,  the small $l$ CMB spectrum is enhanced. This property does not hold for constant DE
equation of state satisfying $w<-1$. In Fig.\ref{cl-cp}, it is shown that the low $l$ CMB spectrum will first be suppressed and then enhanced with the increase
of the strength of the dark scattering. When the scattering is strong enough,
the instability we encountered in the perturbation will cause the blow up in the low $l$ CMB spectrum.
In Fig.\ref{phi-c} we show the behavior of the gravitational potential when
there is dark scattering between DE and DM. The gravitational potential is the source term for the low $l$ CMB spectrum.
For a constant equation of state of DE $w>-1$, a larger value of the scattering results in a further change of the gravitational potential.
When the constant DE equation of state satisfies $w<-1$, there is a sharp increase in the potential as shown in Fig.\ref{phi-cp}
when the scattering is strong enough, which accounts for the blow up in the large scale CMB power.
Since observations suggest the deficit of large scale CMB power, thus we can
rule out the too strong scattering between DE and DM.
Except for the imprints in the low $l$ CMB spectrum, different from the
energy transfer between dark sectors \cite{He.prd83}\cite{Xu:PLB11}, we do not see any
change on the acoustic peaks due to the dark scattering when we choose the DE equation of state to be constant.

\begin{figure}[hptb]
\subfloat[]{
    \includegraphics[width=0.5\textwidth]{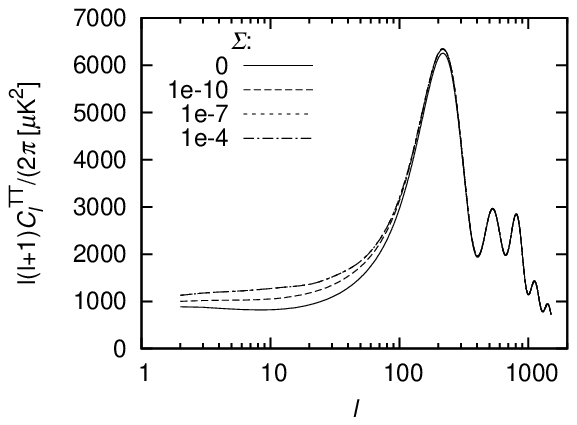}
    \label{cl-dq}
}\\
\subfloat[]{
    \includegraphics[width=0.5\textwidth]{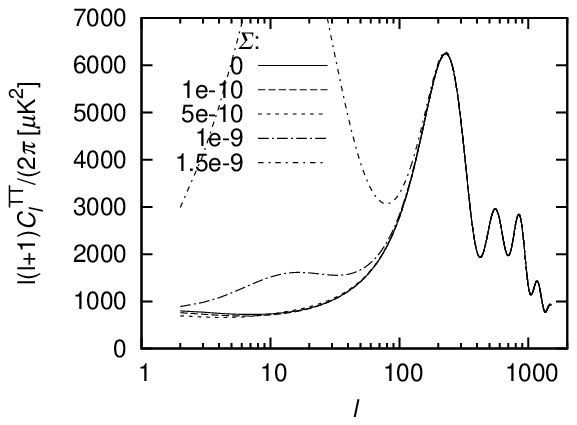}
    \label{cl-dp}
}
\caption{CMB angular power spectrum of models for dynamic DE equation of state in the form $w(a) = w_0a + w_e(1-a)$. In (a) we choose $w_0 = -0.999, w_e = -0.2$ and in (b) we select $w_0 = -1.001, w_e = -1.5$.}
\label{cl-d}
\end{figure}

\begin{figure}[htb]
\subfloat[]{
    \includegraphics[width=0.5\textwidth]{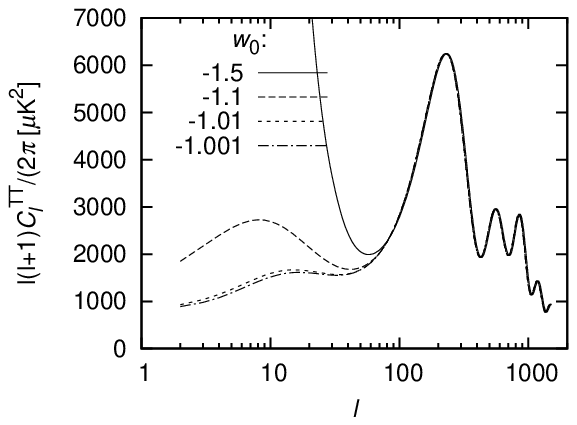}
    \label{cl-w0}
}\\
\subfloat[]{
    \includegraphics[width=0.5\textwidth]{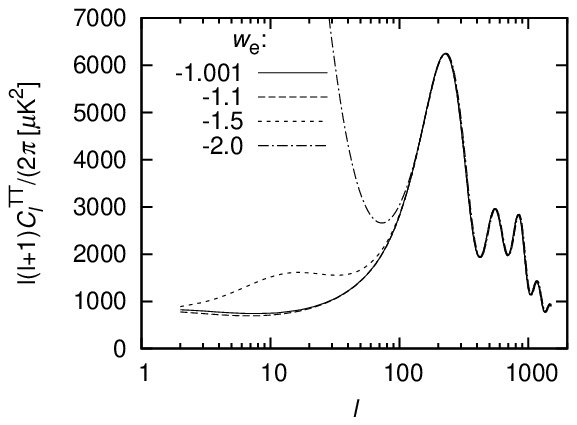}
    \label{cl-we}
}
\caption{CMB angular power spectrum of models with different time-dependant DE equation of state. The DE EoS is in the form $w(a) = w_0a + w_e(1-a)$. $\Sigma = 1 \times 10^{-9} \sigma_T/m_p$.
In (a) we fix $w_e = -1.5$ and in (b) we select $w_0 = -1.001$.}
\label{cl-w}
\end{figure}

Besides the constant equation of state of DE,  we also examine the effect of the dark scattering on CMB when $w$ is time-dependent. We choose the Chevallier-Polarski-Linder (CPL) parametrization\cite{CPL} to describe
the time varying equation of state of DE, where $w$ is expressed as $w(a) = aw_0 + (1-a)w_e$. In the early time,
$a \ll 1$, $w \simeq w_e$; at present, $a \simeq 1$, $w \simeq w_0$. We restrict the equation of state of DE to be in the range either $w>-1$ or $w<-1$ to avoid the singularity at $w=-1$ in the perturbations.

When the DE equation of state is time-dependent and obeys $w>-1$, on the low $l$ CMB spectrum,
we observe a similar effect of the dark scattering to the case for constant DE equation of state
satisfying $w>-1$, see Fig.\ref{cl-dq}.  In addition, the signature of the momentum transfer due to the elastic scattering between dark sectors appears in the first
acoustic peak as well. The change of the acoustic peak mainly appears for a large value of $w_e$, however this effect is washed out when $w_e$ is small.

When the CPL form of the DE equation of state is in the range $w<-1$, the instability can also appear as the case of constant DE equation of state.
In the CPL parametrization, $w_e$ relates to the DE equation of state at the early time, while $w_0$ relates more to the late time DE equation of state. From Fig.\ref{cl-w} we observed that either $w_0$ or $w_e$ deviating significantly from $-1$ will lead to the blow up in the CMB power spectrum at small $l$. The blow up is caused by the quick growth of the velocity when the DE equation of state is much smaller than $-1$. Choosing appropriate DE equation of state, we can obtain similar dependence of the CMB spectrum on the scattering between dark sectors observed for the constant DE equation of state as shown in Fig.\ref{cl-d}. The strength of the dark scattering is constrained to suppress the CMB spectrum at small $l$.

\section{fitting results}

\begin{figure}[htb]
\subfloat[]{
    \includegraphics[width=0.5\textwidth]{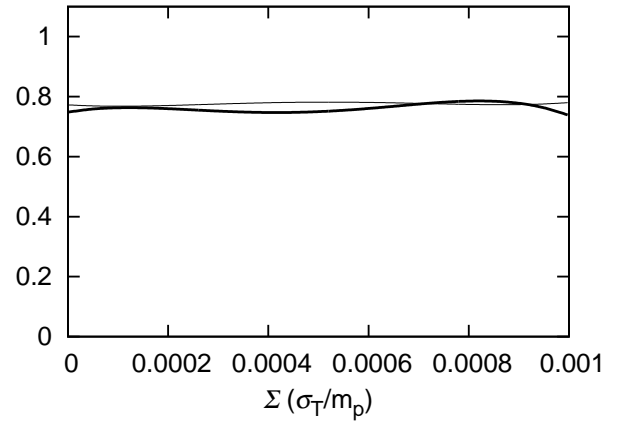}
    \label{sigma-q}
}\\
\subfloat[]{
    \includegraphics[width=0.5\textwidth]{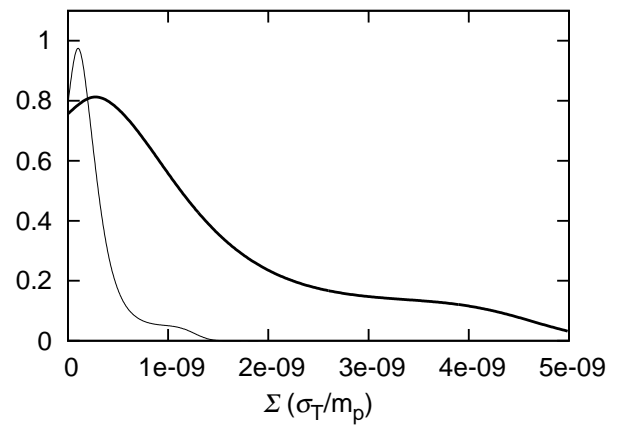}
    \label{sigma-p}
}
\caption{The 1D likelihood of the coupling strength of dark scattering, $\Sigma$. Thin solid lines correspond to fitting results using WMAP data only.
Thick lines correspond to the result of fitting combined data. In (a) we choose $w>-1$ and in (b) we deal with $w<-1$.}
\label{sigma}
\end{figure}

In this section we confront our models with observational data by
implementing joint likelihood analysis. We take the parameter space
as
$$P=(h,\omega_b,\omega_m, \tau, \ln[10^{10}A_s] ,n_s ,w_0, w_e, \Sigma)$$
where $h$ is the hubble constant, $\omega_b=\Omega_bh^2$ and $\omega_{m}=\Omega_{m}h^2$ are baryon and matter abundances,
$\tau$ is the optical depth to last scattering surface,
$A_s$ is the amplitude of the primordial curvature perturbation, $n_s$ is the scalar spectral index. $\Sigma$ here marks the strength of the
dark scattering, which is in the form $\Sigma=\sigma_D/m_c$, where $\sigma_D$ is the cross section of Thomson scattering between DM and DE fluid,
$m_c$ should be the mass of DM particles. Since there is no good approach for computing the DM particle mass, we fit $\Sigma$ from observational data,
instead of the cross section $\sigma_D$. We will choose the CPL parametrization for the DE equation of state
and consider the spatially flat universe with $\Omega_k=0$. To avoid the singularity at $w=-1$ in the perturbation, we concentrate on the range
of the DE equation of state satisfying $w>-1$ and $w<-1$, respectively. Our numerical computation is based on the CMBEASY code \cite{easy}.
We will first fit the CMB anisotropy data from the seven-year Wilkinson Microwave Anisotropy Probe(WMAP).
The fitting results are listed in Table.\ref{bestfit}. The coupling strength $\Sigma$ is in the unit of $\sigma_T / m_p$ where $\sigma_T$ is the cross section of Thomson scattering and $m_p$ is proton mass.
The 1d marginalized likelihood for the strength of the dark scattering is shown in the thin solid lines in Fig.\ref{sigma} when
we only use the CMB anisotropy data from the seven-year WMAP observation.

\begin{table*}[htbp]
\caption{The cosmological parameters from the global fitting.} \centering \label{bestfit}
\renewcommand\arraystretch{1.5}
\begin{tabular}{ccccc}
    \toprule
    & \multicolumn{2}{c}{WMAP} & \multicolumn{2}{c}{WMAP+BAO+SN} \\
    \cline{2-5}
    Parameter & $w>-1$ & $w<-1$ & $w>-1$ & $w<-1$ \\
    \hline
    $\Omega_m h^2$ & $0.126^{+0.006}_{-0.006}$ & $0.134^{+0.005}_{-0.005}$ & $0.131^{+0.004}_{-0.004}$ & $0.136^{+0.004}_{-0.004}$ \\
    $\Omega_b h^2$ & $0.0231^{+0.0008}_{-0.0007}$ & $0.0220^{+0.0006}_{-0.0006}$ & $0.0228^{+0.0007}_{-0.0006}$ & $0.0220^{+0.0005}_{-0.0005}$ \\
    $h$ & $0.724^{+0.027}_{-0.027}$ & $0.752^{+0.074}_{-0.031}$ & $ 0.706^{+0.013}_{-0.013}$ & $0.698^{+0.013}_{-0.012}$ \\
    $\tau$ & $0.089^{+0.016}_{-0.015}$ & $0.083^{+0.014}_{-0.013}$ & $ 0.084^{+0.015}_{-0.014}$ & $0.082^{+0.014}_{-0.013}$ \\
    $n$ & $0.992^{+0.024}_{-0.021}$ & $0.955^{+0.014}_{-0.014}$ & $ 0.982^{+0.019}_{-0.017}$ & $0.955^{+0.013}_{-0.013}$ \\
    ${\rm ln} [10^{10}A_s]$ & $3.067^{+0.036}_{-0.035}$ & $3.073^{+0.034}_{-0.034}$ & $ 3.070^{+0.035}_{-0.033}$ & $3.076^{+0.033}_{-0.032}$ \\
    $w_0$ & $<-0.948$ & $>-1.260$ & $<-0.950$ &$>-1.031$ \\
    $w_e$ & $<-0.943$ & $>-1.370$ & $<-0.960$ & $>-1.077$ \\
    $\Sigma$ & N/A & $ <3.295 \times 10^{-10}$ & N/A & $<1.665 \times 10^{-9}$ \\
    \lasthline
\end{tabular}
\end{table*}

In order to get tighter constraint on the parameters, we employ BAO and SNIa data sets in addition to WMAP.
The BAO distance measurements \cite{BAO} which are obtained from analyzing clusters of galaxies and test a different region in the sky as
compared to CMB. BAO measurements provide a robust constraint on the
distance ratio
\begin{equation}
d_z = r_s(z_d)/D_v(z)
\end{equation}
where $D_v(z)\equiv[(1+z)^2D_A^2z/H(z)]^{1/3}$ is the effective distance \cite{Eisenstein}, $D_A$ is the angular diameter distance,
and $H(z)$ is the Hubble parameter. $r_s(z_d)$ is the comoving sound horizon at the baryon drag epoch where the baryons decoupled from photons.
We numerically find $z_d$ using the condition $\int_{\tau_d}^{\tau_0}\dot{\tau}/R=1,R=\frac{3}{4}\frac{\rho_b}{\rho_{\gamma}}$
as defined in \cite{wayne}. The $\chi^2_{BAO}$ is calculated as \cite{BAO},
\begin{equation}
\chi^2_{BAO}=(\vec{\textbf{d}}-\vec{\textbf{d}}^{obs})^T\textbf{C}^{-1}(\vec{\textbf{d}}-\vec{\textbf{d}}^{obs})
\end{equation}
where $\vec{\textbf{d}}=(d_{z=0.2},d_{z=0.35})^T$, $\vec{\textbf{d}}^{obs}=(0.1905,0.1097)^T$ and the inverse of covariance matrix reads
\cite{BAO}
\begin{equation}
\textbf{C}^{-1}=\left(
                  \begin{array}{cc}
                    30124 & -17227\\
                    -17227 & 86977 \\
                  \end{array}
                \right).
\end{equation}
Furthermore, we add the BAO A parameter \cite{Eisenstein},
\begin{eqnarray}
A&=&\frac{\sqrt{\Omega_m}}{E(0.35)^{1/3}}\left[\frac{1}{0.35}\int_0^{0.35}\frac{dz}{E(z)}\right]^{2/3}\nonumber \\
 &=&0.469(n_s/0.98)^{-0.35}\pm 0.017
\end{eqnarray}
where $E(z)=\frac{H(z)}{H_0}$ and $n_s$ are the scalar spectral index.

We also use the compilation of 397 Constitution samples from supernovae survey \cite{SNeIa397}. We compute
\begin{equation}
\chi^2_{SN}=\sum \frac{[\mu(z_i)-\mu_{obs}(z_i)]^2}{\sigma_i^2}\quad ,
\end{equation}
and marginalize the nuisance parameter.

We implement the joint likelihood analysis,
\begin{equation}
\chi^2=\chi^2_{WMAP}+\chi^2_{SN}+\chi^2_{BAO}.
\end{equation}

From Fig.\ref{sigma-q}, we see that CMB
anisotropy probes can not put tight constraints
on the intensity of dark scattering when $w>-1$.
One reason to understand this is that the dark
scattering mainly influences the small $l$ CMB
power spectrum through the late ISW effect, while
the CMB data at small $l$ are poor and lacking.
Further, from the theoretical study we learnt
that the elastic scattering between DM and DE is
suppressed when the equation of state of DE is
close to $-1$. From the fitting results for the
equation of state of DE, it is really close to
$-1$ in the recent epoch. $w\sim -1$ is further
constrained by adding BAO or SNIa data. When the
DE equation of state is in the range $w<-1$, we
see from Fig.\ref{sigma-p} that the dark
scattering is tightly constrained. This is
because when $w<-1$ the small $l$ CMB spectrum is
enhanced when the scattering is strengthened and
blows up if the scattering is strong enough. The
observed suppression of the CMB data at low $l$
strongly puts the upper bound on the strength of
the scattering. For this reason, a tight
constraint is put on the dark scattering when the
DE equation of state obeys $w<-1$. Including the
BAO or SNIa data, the equation of state of DE is
constrained further close to $-1$, the cross
section $\sigma_D$ can then be arbitrarily large.
All we can really constrain is
$\Sigma'=(1+w)\sigma_D/m_c$. There is a
degeneracy between $(1+w), \sigma_D$ and $m_c$.
When $w\sim -1$ is constrained at present, the
constraints on $\sigma_D$ will be poor.

If we assume $ m_c = m_p$, where $m_p$ is the
mass of proton, from the fitted value of $\Sigma$
in Table.1, we can estimate  $\sigma_D < 3.295
\times 10^{-10} \sigma_T$ (for WMAP alone) and
$\sigma_D < 1.665 \times 10^{-9} \sigma_T$ (from
the combined constraint).

To conclude, when $w>-1$, the effect of elastic scattering between dark sectors has little influence on the CMB power spectrum, which is different from the energy transfer discussed in our previous papers. Thus the CMB observation cannot put tight constraint on the dark scattering when DE equation of state $w>-1$. When $w<-1$, in the constrained ranges of $w_0$ and $w_e$, big strength of the dark scattering will lead to the sharp blow up of CMB spectrum at small $l$. This effect is similar to the energy transfer discussed in \cite{He.prd83}\cite{He.prd80}\cite{Xu:PLB11}. Using the precise CMB data, the strength of the dark scattering can be constrained tightly.

\section{Conclusion and Discussion}
In this paper we generalized the study of the interaction between DE and DM.
Instead of considering the energy exchange between DE and DM in the literatures, in this work we
investigated the momentum transfer by exploring the elastic scattering between dark sectors. Adopting the
phenomenological description of the DM elastically scattering within the DE fluid \cite{Simpson}, which is analogous to the
Thomson scattering for baryons and photons, we examined the impact of the dark scattering on CMB observations.
We found that in addition to the growth of structure \cite{Simpson},  the elastic scattering between the DM and DE fluid can also leave
cosmological signals in CMB. The imprint of the dark scattering in CMB highly depends on the equation of state of DE. When the DE equation of state
$w<-1$, we can constrain the dark scattering well from CMB observations. However when $w>-1$, the cosmological signal of the scattering is weak.
The impact of the dark scattering is suppressed when $w\sim -1$.

From our result, we learnt that the CMB probes are not sufficient to constrain the dark scattering. It would be interesting to carry out the complementary constraints by combining the CMB test with the large scale structure tests as employed in the study of
the energy transfer between dark sectors \cite{He.jcap07}-\cite{Baldi.MNRAS11}. Progress in this direction will be reported in the future.

\acknowledgements{This work has been supported
partially by the National Basic Research Program
of China under grant 2010CB833000 and NNSF of
China No. 10878001. EA wishes to thank FAPESP and
CNPq (Brazil) for financial support. BW would
also like to acknowledge the Shanghai Science and
Technology Commission for the grant 11DZ2260700.}

\end{document}